\documentclass[aps,pre,showpacs,twocolumn]{revtex4}
\usepackage{epsfig}
\usepackage{color}
\usepackage{dcolumn,amsmath,amsthm,amscd,amsfonts,amssymb,graphics,graphicx,eucal}

\begin{document}


\title{Synchrotron radiation of dissipative solitons in optical fiber cavities}

\author{A. V. Yulin}
\address{
ITMO University 197101, Kronverksky pr. 49, St. Petersburg, Russian Federation
}
\begin{abstract}
New resonant emission of dispersive waves by oscillating solitary structures is considered analytically and numerically.The resonance condition for the radiation 
is derived and it is demonstrated that the predicted resonances match the spectral lines observed in numerical simulations perfectly. The complex recoil of the 
radiation on the soliton dynamics is  discussed.
\end{abstract}

\maketitle

Since the pioneering work by Akhmediev and Karlsson \cite{AkhmedievKarlsson} the resonant radiation of optical solitons propagating in fibers with high order dispersion has been attracting much of attention because of both the fundamental interest and practical importance \cite{Skryabin_rev,Genty_rev}. The condition of the resonance is the equality of the soliton velocity to the phase velocity of a dispersive wave and in this way the radiation is similar to the effect of Cherenkov radiation of charged particles. A similar effect of transitional radiation of Bragg solitons was later discovered in periodic media \cite{yulin0}.

The aim of the present Letter is to address the resonant radiation of oscillating optical solitons and to reveal the analogy of this radiation to the synchrotron radiation of charges appearing when the velocity of relativistic charges varies periodically in time \cite{synchrotron}.  

It should be noted here that from the point of view of the field theory the synchrotron radiation can be interpreted as an excitation of resonant waves by an oscillating localized source moving at a relativistic velocity. The periodical variations of the velocity of a relativistic charge is just a way to obtain the right driving force in the equation for the electromagnetic waves. 

Aiming to develop the analogy between synchrotron radiations of the charges and solitons it is much simpler and more instructive to consider one-dimensional oscillating solitons  rather than two-dimensional nonlinear localized structures relativistically moving along one direction and wobbling in the transverse 
direction (though the latter is also possible).

The effects of Cherenkov and synchrotron radiation are widely used in vacuum electronics, in particular in forward- and backward waves oscillators. The optical counterparts of Cherenkov effects can play an important role in the generation of new optical frequencies by the solitons propagating in fibers with high order 
dispersion. Indeed, since 1995 when Cherenkov radiation of optical solitons was reported for the first time this effect has been actively studied both experimentally and theoretically, especially in the context of the generation of optical supercontinuum  \cite{Skryabin_rev,Genty_rev}.  Recently a number of papers 
were published on Cherenkov emission shed by shock waves \cite{Conforti} and ultra-short pulses \cite{Rubino,Biancalana2,Conforti2}. 

All these effects are important for the explanation of the dynamics of solitons and domain walls in wave-guiding systems. The effect of synchrotron emission of oscillating solitons can also affect the dynamics of the solitons strongly and thus is of interest from both fundamental and practical points of view. 

Optical solitons in fibers seem to be the best candidates for experimental observation of the discussed effect of synchrotron emission, however similar effects can take place in polariton condensates, 
plasmas, hydrodynamics and other systems of different physical origins.

The resonant emission is possible not only in the conservative but also in weakly dissipative systems. In this Letter we consider the radiation in annular dissipative optical cavities with external pump. The advantage of dissipative systems is that the resonant radiation can occur there in a stationary regime when it is not disguised by any transitional processes. 

Our choice is also motivated by the fact that these systems are not only a convenient test bench for investigation of synchrotron radiation but are of interest on their own and  have been actively investigated in recent years theoretically and experimentally\cite{Leo,Chembo,Kippenberg,DelHaye,Matsko,Herr,Tlidi,Mussot,Leo2,Coen,Lamont,Gelens,Parra}.

Propagation of oscillating solitons in the presence of high order dispersion was studied experimentally and numerically and very reach dynamics of the solitons  was reported  \cite{Leo3}. Cherenkov radiation of oscillating cavity solitons was recently considered in \cite{MilianSkryabin} but the effect of the synchrotron 
radiation was out of the scope of that paper. In the present Letter we focus on synchrotron radiations specific for oscillating solitons. 

To describe the cavity we adopt a well known Lugiato-Lefever model in the form as it is written in \cite{Gelens}
\begin{eqnarray}
\partial_t A=-(1+i\theta)A+i|A|^2A+i\partial_x^2A+d_3\partial_x^3 A+P
\label{Lugiato}
\end{eqnarray}
where $\theta$ is the detuning between the cavity resonance and the frequency of the pump, $P$ is the amplitude of the pump and $d_3$ is the third order 
dispersion. To make the resonance spectral lines well pronounced we choose the detuning to be equal to a realistic value $\theta=15$ which is just slightly greater then the values used in \cite{Gelens}. 
The boundary conditions for annular cavities are periodic but it is checked that for the for large cavities considered in the present Letter the boundary conditions do not affect the dynamics of the solitons. 

Equation (\ref{Lugiato}) has bright soliton solutions that loose stability through Andronov-Hopf bifurcation leading to the formation of an oscillating soliton \cite{Firth1,Firth2} for the pumps exceeding a threshold value. For the chosen parameters the threshold pump is $P_{th}\approx 7$. Aiming to consider the behaviour of oscillating solitons we choose the pump above the threshold $P=8$ and consider dynamics of the soliton.

\begin{figure}
\includegraphics[width=\columnwidth]{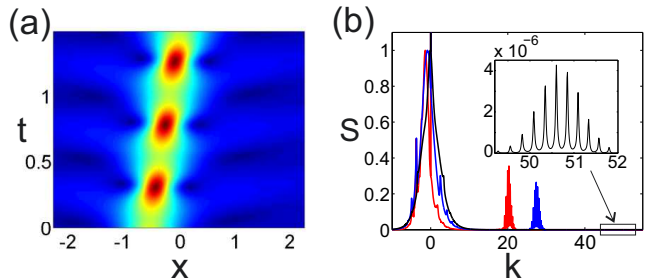}%
\caption{(Color online) Panel (a) shows the temporal evolution of the field of an oscillating soliton for  $d_3=0.02$ and $P=8$. Typical spectra of the fields are shown in panel (b) for $d_3=0.02$, $d_3=0.04$ and 
$d_3=0.06$ by the black, blue and red lines. The synchrotron radiation spectrum for $d_3=0.02$ is shown in the inset.}%
\label{fig1}
\end{figure}

The stationary temporal evolution of the intensity of the field is shown in panel (a) of Fig.~\ref{fig1} for relatively small value of third order dispersion $d_3=0.02$. It is clearly seen that the soliton oscillates in time. The typical spectra of the field are shown in panel (b). The resonant emission is very weak for $d_3=0.02$ but still clearly seen at $k \approx 50$ as predicted by the resonance condition \cite{MilianSkryabin}. Much stronger radiation is seen at $k \approx 28$ and $k \approx 20$ for $d_3=0.04$ and $d_3=0.06$.

A very important fact is that the spectrum of the resonant radiation consists not of a single line as it should be in the case of Cherenkov radiation but of a series of well resolved peaks. Below we will show that, indeed, one of these lines corresponds to Cherenkov radiation whereas the other lines can be 
refereed as synchrotron radiation. 

It is worth mentioning here that narrow radiation lines at low $k$ overlapping with the soliton spectrum are also seen in panel (b). These lines can be understood as non-relativistic radiation of the oscillating solitons  and thus are analogous to, for example, cyclotron radiation of charged particles. 

It is worth mentioning here that the equation (\ref{Lugiato}) is of course not invariant in respect of Lorentz transformation and in this Letter we use the term "relativistic radiation" only in the sense that the velocity of the soliton is close to the phase velocity of the dispersive waves. 

To understand the observed phenomenon let us derive the conditions of resonant radiation of oscillating solitons. Here we only sketch the derivation, the complete analysis will be presented elsewhere. 

Assuming that the radiation is weak it is possible to look for a solution in a form $A=A_0 +u[t, x]$ where $A_0$ describes the soliton oscillating with period $T$  and moving with the velocity $v$; the term $u$ accounts for the small correction. Here and below the arguments of the functions are given in square brackets. 

It is convenient to write an equation for $u$ in vector form $\vec u= (Re \, u, Im \, u)^T$
\begin{eqnarray}
\partial_t \vec u+ \hat L \vec u = \vec f .
\label{correction}
\end{eqnarray}
The linear operator $\hat L$ has the form
$$\hat L = \left(
                  \begin{array}{cc}
                    1+Im \, A_0^2 +d_3\partial_x^3& 2I-Re \, A_0^2 -\theta+\partial_x^2\\
                    -2I -Re \, A_0^2+\theta-\partial_x^2 & 1-Im \, A_0^2 +d_3\partial_x^3\\
                  \end{array}
                \right),
$$
where $I=|A_0|^2$. The right hand side is given by $\vec f= (Re \, f_0, Im \, f_0)^T$ with
\begin{eqnarray}
f_0=P-(1+i\theta-i|A_0|^2-i\partial_x^2 -d_3 \partial_x^3) A_0-\partial_t A_0.
\label{rhs1}
\end{eqnarray}

The left hand side of equation (\ref{correction}) describes propagation of waves on the background hosting the soliton. In the conservative case the resonance of the right hand side $\vec f$ with a delocalized eigenmode of the medium results in the formation of a continuous wave propagating away from the soliton.  In the dissipative case the length of the radiation tail generated by the resonance is, of course, always finite but can be very large for low losses.

The oscillations of the soliton make the coefficients in $\hat L$ to be periodic functions of time. It means that the eigenfunctions of the operator $\hat L$ are Bloch functions in time. The important fact, however, is that far from the soliton the problem becomes uniform in both space and time and thus the asymptotic of the delocalized  Bloch modes $\xi_{\omega}$ at $x \rightarrow \pm \infty$ are just plane waves 
$\xi_{\omega} \sim e^{i\omega t - ikx}$ parametrized by the frequency $\omega$ and the wave vector $k$. So the
dispersion characteristics if the delocalized eigenmodes is nothing else but the dispersion characteristics of plane waves $\omega=\omega_{\pm}[k]$ on the background  
\begin{eqnarray}
\omega_{\pm}=i+d_3k_r^3 \pm\sqrt{(\theta+k^2-I_0)(\theta+k^2-3I_0)},
\label{disp}
\end{eqnarray}
where $I_0=I[\pm\infty]$ is the intensity of the background.

To proceed let us represent the right hand side of (\ref{correction}) in the form $\vec f =\sum_l \vec U_l[x-vt] \cdot e^{il\omega_0 t}$ where $\omega_0=2\pi/T$ is the fundamental frequency of the soliton. Then each of the term $U_l$ in the sum can be represented as series over the eigenfunctions of the operator $\hat L$. From the fact that all functions $U_l[x-vt]$ are moving without changing their shapes it follows that the frequencies of the driving force components depend on their wave vectors as $\omega=kv+l\omega_0$.

The emission of radiation takes place when a harmonic of the right hand side $\vec f$ is in resonance with an eigenwave of the medium. This gives the condition of the phase synchronism
\begin{eqnarray}
\omega_{\pm}(k_r)=k_rv+l \omega_0.
\label{resonance_condition}
\end{eqnarray}
For $l=0$ this is the condition of Cherenkov radiation. If the condition is satisfied for $l \neq 0$ then the radiated wave is excited by $l$-th temporal harmonic of the oscillating soliton.

\begin{figure}
\includegraphics[width=\columnwidth]{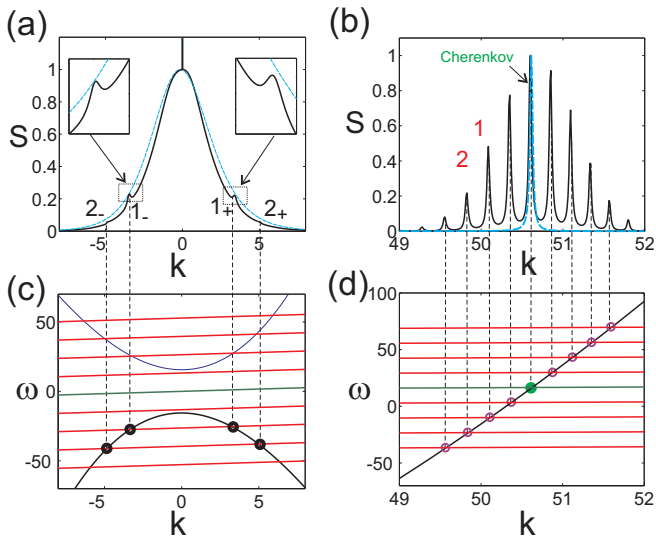}%
\caption{(Color online) Panel (a) shows the spectrum at low $k$ of the oscillating (thicker black line) and non-oscillating solitons (thinner blue line). 
The spectrum in the vicinity of Cherenkov resonance is shown in panel (b). The graphical solutions of the resonance conditions (\ref{resonance_condition}) 
are shown in panels (c) and d), the green line corresponds to the Cherenkov resonance and the red ones to the resonances with $l \neq 0$. Resonances 
marked as $1_+$, $1_-$ on panel (a) and by $1$ on panel (b) are given by different crossings of the same red line with the dispersion characteristic 
of the linear excitations on the soliton background; the same is for the resonances marked by $2$. The third order dispersion is $d_3=0.02$, the pump 
is $P=8$.}%
 \label{fig2}
\end{figure}

Let us now discuss how the derived resonance conditions explain the results of the numerical experiments. 
To do this we study the spectrum and the resonance conditions at low $k$ and at $k$ close to Cherenkov resonance. We start with the resonances at small wave vectors $k$. The spectrum obtained in numerical simulations is shown in panel (a) of Fig.~\ref{fig2}. The spectrum looks like a bell-shaped function with the narrow line at $k=0$ corresponding to the background. Comparing spectra of non-oscillating and oscillating solitons one can notice that in the latter case several additional maxima appears on the spectral curve.

The graphical solutions of the resonance condition (\ref{resonance_condition}) are shown in panel (c). The soliton velocity and the frequency of the soliton was extracted from numerical simulations. It is seen for $|l| \geq 2$ the resonance condition is met. Comparing the positions of the crossings with the positions of the small local maxima of the spectrum we see that they match perfectly. The phase velocity of these resonant modes are much higher comparing to the soliton velocity and thus developing the analogy between the radiation of solitons and moving charges we can refer these resonances as cyclotron radiation of solitons. 

\begin{figure}
\includegraphics[width=\columnwidth]{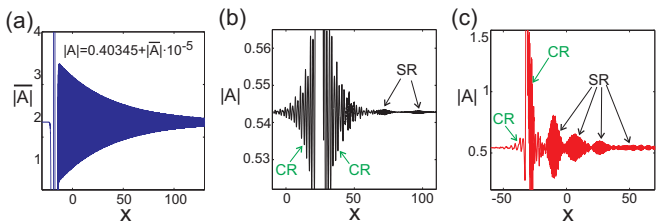}%
\caption{(Color online) Radiation field of the non-oscillating soliton is shown in panel (a), $P=6$  $d_3=0.02$.  The radiations of the oscillating solitons 
($P=8$) are illustrated in panels (b) and (c) for $d_3=0.02$ and $d_3=0.06$. The synchrotron and cyclotron radiations are marked by "SR" and "CR" correspondingly. }%
\label{fig4}
\end{figure}

It is worth noticing that the emitted waves with $k<0$ propagate with the group velocity greater than the velocity of the soliton and so the radiation appears in front of the soliton. The emitted radiation with $k>0$ propagates in the opposite direction and so the radiation appears behind the soliton. Topologically it is obvious that if there is a resonance at a positive $k$ then there is a resonance at a negative $k$ and so cyclotron radiation always appears simultaneously in front and behind the soliton, see panels (b) and (c) of  Fig.~\ref{fig4} showing the field distributions of the oscillating solitons. 

Now let us consider the radiation with the frequencies close to the Cherenkov resonance.  
The crossing of the green line ($l = 0$) with the dispersion characteristics in panel (b) gives 
the position of Cherenkov resonance matching one of the spectral lines observed in the 
numerical experiment perfectly. The other resonances ($l \neq 0$) fit the other spectral lines. So we can explain the spectral structure of soliton radiation as a number of lines produced by different temporal harmonics of the oscillating soliton. 

It is obvious that the discussed waves have phase velocities close to the velocity of the soliton 
(for Cherenkov resonance they are exactly equal) and thus these radiations can be interpreted as 
relativistic radiation of the oscillating soliton, ie synchrotron radiation. In the considered 
case Cherenkov and synchrotron radiations always have group velocities higher than the velocity 
of the soliton and so the radiation must appear in front of the soliton.  

It is instructive to compare the radiation of oscillating and not oscillating solitons. The blue curves in panels (a) and (b) show the spectrum of a non oscillating soliton forming in the case of a weaker pump $P=6$. The velocities of the solitons are practically the same for $P=6$ and $P=8$ and so 
the Cherenkov spectral lines of the oscillating and non-oscillating solitons practically coincide. However it is clearly seen that the non-oscillating soliton emits neither cyclotron no synchrotron radiation.

In the case of third order dispersion the synchrotron and Cherenkov resonances always appear together, but 
it is important to  remark that there are dispersion characteristics allowing for synchrotron resonances in the absence of the Cherenkov one.  

To shed more light on the resonant emission we consider the shapes of the radiation fields. Panel (a) of Fig.~\ref{fig4} shows stationary Cherenkov radiation of non-oscillating soliton pumped by $P=6$ when third order dispersion is equal to $d_3=0.02$. The radiation contains only one mode, see the spectrum in panel (b) of Fig.~\ref{fig2}, and decays exponentially with the rate $1/|v_{g}-v|$.

For stronger pump $P=8$ the soliton develops oscillations and, as discussed above, the cyclotron radiation appears in front and behind the soliton. This radiation is clearly seen in panel (b). The radiations having small group velocity decays along $x$ quickly, so that this radiation is seen only in the vicinity of the soliton. 

Cherenkov and synchrotron radiations with large $k$ have bigger group velocities and so decay slower then cyclotron radiation. This radiation marked by SR is weak and barely visible for $d_3=0.02$ but becomes much stronger for $d_3=0.06$, see panels (b) and (c) 

A very important fact is that the radiation of an oscillating soliton is pulsating whereas the radiation of a non-oscillating soliton is a monotonically decaying continuous wave. The pulsations can be explained by the interference of synchrotron emissions at close frequencies. The distances between the synchrotron resonances can be estimated as $\Delta k=\omega_0/|v_{gs}-v|$ where $v_{gs}=\partial_k \omega(k_r)$ is the group 
velocity of the radiation. Then the length of the radiation pulses is $2\pi v_{gs}/\Delta \omega$  giving a good estimate of $\approx 15$ for $d_3=0.06$.

\begin{figure}
\includegraphics[width=\columnwidth]{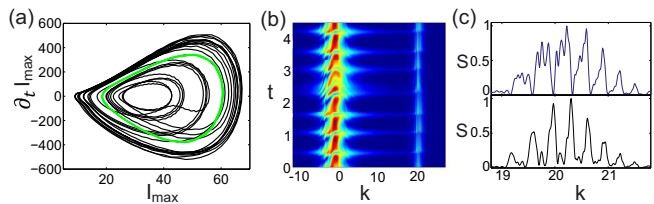}%
\caption{(Color online) Black and green lines in panel (a) show the dynamics of the solitons in $\partial_t I_{max}$-$I_{max}$ plane for  $d_3=0.06$ and $d_3=0.02$ correspondingly. The evolution of the spectrum of an oscillating soliton is shown in panel (b) for $d_3=0.06$, panel (c) shows the spectra of synchrotron radiation at two different times. The pump is $P=8$.}%
\label{fig3}
\end{figure}

Now we turn our attention to a recoil from the radiation on the oscillating solitons. From numerics we extract the dependence of maximum intensity of the soliton $I_{max}[t]$ on time and calculate its temporal derivative $\partial_t I_{max}[t]$. These dependencies parametrically define the curves in $I_{max}[t]$-$\partial_t I_{max}[t]$ plane, see (a) of Fig.~\ref{fig3}. For a periodic motion this curve is a closed loop and this is the case so for small $d_3=0.02$. However for large $d_3=0.06$ the motion looks very complicated. The temporal evolution of the spectrum is also not periodic for $d_3=0.06$, see panels (b) and (c).

This behaviour can be explained by the fact that the pulsations of the synchrotron radiation causes 
pulsating recoil changing the intensity and velocity of the soliton. The variations of 
the soliton parameters in their turn affect the instantaneous frequencies of the radiation.  For large $d_3$ the interplay of these effects leads to a complex dynamics of the solitons.  
Possibly this explains the complicated dynamics of cavity solitons discussed in \cite{Leo3}.

Now we summarize the main results of the paper. It is shown that oscillating solitons can emit resonant radiation analogous to the synchrotron and cyclotron radiations of moving charges. The resonance condition for the radiation is derived and it is demonstrated that the predicted resonances match the 
observed spectral lines very precisely. The structure of the emitted radiation and the effect of the recoil of the radiation on the solitons are discussed. The reported results are of general physical nature and thus can be observed not only in fiber cavities but in a wide class of physical systems supporting oscillating solitons.

The author thanks Prof. D. Skryabin and Dr. R. Driben for useful discussions.
This work was financially supported by the Government of the Russian Federation (Grant 074-U01) through ITMO Early Career Fellowship scheme.


\begin{thebibliography}{99}

\bibitem{AkhmedievKarlsson} N. Akhmediev and M. Karlsson,  Phys. Rev. A {\bf 51}, 2602, (1995)
 
\bibitem{Skryabin_rev} D. V. Skryabin and A. V. Gorbach, Rev. Mod. Phys. 82, 1287, (2010)

\bibitem{Genty_rev} J. M. Dudley, G. Genty, and S. Coen, Rev. Mod. Phys. 78, 1135, (2006)

\bibitem{yulin0} A.V. Yulin, D.V. Skryabin, and P.St.J. Russel, Phys. Rev. Lett. {\bf 91}, 260402, (2003)


\bibitem{synchrotron} F.R. Elder et al., Physical Review, {\bf 71 }, 11, 829,  (1947)


\bibitem{Conforti} M. Conforti, F. Baronio, and S. Trillo, Phys. Rev. A {\bf 89},  013807 (2014)

\bibitem{Rubino} E. Rubino et al., Phys. Rev. Lett. {\bf 108}, 253901 (2012)

\bibitem{Biancalana2} F. Biancalana, Physics 5, 68 (2012)


\bibitem{Conforti2} M. Conforti et al., Optics Express {\bf 21}, No. 25, 31239, (2014)


\bibitem{Skryabin_Scie} D. V. Skryabin et al., Science, {\bf 301}, 1705 (2003)

\bibitem{Biancalana} F. Biancalana, D. V. Skryabin, and A. V. Yulin, Phys. Rev. E, {\bf 70}, 016615 (2004)


\bibitem{Leo} F. Leo et al.,  Nature Photonics 4, 471 (2010)



\bibitem{Chembo} Y. K. Chembo and N. Yu, Phys. Rev. A 82, 033801 (2010)


\bibitem{Kippenberg} T.J. Kippenberg, R. Holzwarth, S.A. Diddams, Science 332, 555–559 (2011)

\bibitem{DelHaye} P. Del'Haye et al., Phys. Rev. Lett. 107, 063901 (2011)

\bibitem{Matsko} A. B. Matsko et al.,  Opt. Lett. 36, 2845 (2011)

\bibitem{Herr}  T. Herr et al., Nature Photonics, {\bf 8}, 145 (2014)


\bibitem{Tlidi} M. Tlidi et al., Optics Lett. {\bf 32}, No. 6, 662, 2007

\bibitem{Mussot} A. Mussot et. al, Phys.Rev.Lett. {\bf 101}, 113904 (2008)

\bibitem{Leo2} F. Leo et al., Phys. Rev. Lett. {\bf 110}, 104103 (2013)

\bibitem{Coen} S. Coen et al., Opt. Lett. 38, 37–39 (2013)

\bibitem{Lamont} M. R. E. Lamont, Y. Okawachi, and A. L. Gaeta, Opt. Lett. 38, 3478–3481 (2013)

\bibitem{Gelens} L. Gelens et al.,  Optics Letters, Vol. 39, Issue 10, 2971,  (2014)

\bibitem{Parra} P. Parra-Rivas, D. Gomila, M.A. Matias, S. Coen, L. Gelens, Phys. Rev. A {\bf 89}, 043813 (2014)



\bibitem{Leo3} F. Leo et al., Optics Express, {\bf 21}, Issue 7, 9180 (2013)

\bibitem{MilianSkryabin} C. Milian, and D.V. Skryabin, Optics Express {\bf 22}, No. 3, 3732 (2014)


\bibitem{Firth1} W.J. Firth, A. Lord, and A.J. Scroggie, Phys. Scr., {\bf 67}, 12, (1996)

\bibitem{Firth2} W.J. Firth et al., J. Opt. Soc. Am. B, {\bf 19}, 747 (2002).


\end{thebibliography}
\end{document}